\documentclass[np2,twoside,twocolumn,showkeys,nofootinbib]{revtex4-2}

\usepackage{amssymb}
\usepackage{amsmath}
\usepackage{graphicx}
\usepackage{bm,multirow}
\usepackage{amsmath}
\usepackage{verbatim}
\usepackage{hyperref}
\usepackage{subfigure}

\newcommand{\etal}{{\it et al.}}
\newcommand{\be}{\begin{equation}}
\newcommand{\ee}{\end{equation}}
\newcommand{\msun}{M_{\odot}}

\volumeyear{Month Year} 
\startpage{1}
\endpage{3}

\begin{document}

\title[$\ll$ Review Article $\gg$ Guideline for NPS articles]{Impact of Higher-order Tidal Corrections on the Measurement Accuracy of Neutron Star Tidal Deformability}

\author{Gyeongbin \surname{Park}}
\affiliation{Department of Physics, Pusan National University, Busan 46241, Korea}

\author{Hee-Suk \surname{Cho}}
\email{chohs1439@pusan.ac.kr} 
\affiliation{Department of Physics, Pusan National University, Busan 46241, Korea}
\affiliation{Extreme Physics Institute, Pusan National University, Busan 46241, Korea}

\author{Chang-Hwan \surname{Lee}}
\affiliation{Department of Physics, Pusan National University, Busan 46241, Korea}

\begin{abstract}

Gravitational waves emitted by binary neutron stars (BNS) provide information about the internal structure of neutron stars (NSs), helping to verify dense matter equations of state.
We investigate how the measurement accuracy of NS's tidal deformability can be improved by incorporating the higher-order post-Newtonian (pN) tidal corrections up to 7.5 pN.
We assume an aligned-spin BNS system and adopt TaylorF2, which is the most commonly used pN waveform model.
To calculate the measurement error, we use a semi-analytic method, Fisher Matrix, which is much faster than performing parameter estimation simulations.
We employ Universal Relation to remove additional parameters that appear in higher-order corrections beyond 6 pN.
We find that the effect of tidal corrections shows no behavior of convergence with increasing pN orders.
Assuming a fiducial binary NS system whose physical parameters are compatible with GW170817,
we find that the measurement error of tidal deformability ($\tilde{\lambda}$) decreases linearly as the effective spin ($\chi_{\rm eff}$) increases
and the tidal deformability can be better measured for stiffer equation of states.

\end{abstract}


\maketitle

\section{INTRODUCTION}
\label{sec:introduction}

In 2017, the observation of GW170817 \cite{LIGOScientific:2017vwq, LIGOScientific:2018hze} and its electromagnetic counterpart \cite{LIGOScientific:2017zic, Goldstein:2017mmi, Savchenko:2017ffs} originated from a binary neutron star (BNS) coalescence opened a new era of multi-messenger astrophysics.
Because gravitational waves (GWs) from a BNS system have information about the internal structure of neutron stars (NSs), such observation has led to advances in nuclear physics as well as astrophysics.
Especially, this GW signal could provide tighter constraints on the maximum mass \cite{Margalit_2017, Rezzolla:2017aly, PhysRevD.97.021501, PhysRevD.96.123012}, radii \cite{PhysRevLett.121.161101, PhysRevLett.120.261103, Raithel:2018ncd}, speed of sound \cite{EPJ252.2021}, and tidal deformability \cite{PhysRevLett.121.161101, PhysRevLett.120.172703, PhysRevLett.120.261103} of NSs.
Moreover, those observations can help us to understand the fundamental interaction of extreme dense matter by verifying the equation of states (EOS) \cite{LIGOScientific:2018cki, PhysRevX.9.011001} and to test the theory of the gravitation \cite{PhysRevD.103.122002, PhysRevD.107.044020, LIGOScientific:2017zic, PhysRevLett.116.221101}.
The future 3 generation (3G) detectors such as Einstein Telescope (ET) \cite{Punturo_2010, Hild_2011} and Cosmic Explorer (CE) \cite{Abbott_2017}
will make more GW detections and generate more accurate parameter estimation.

In a compact binary system containing a NS, the NS can be deformed by tidal fields produced by its companion star. 
This finite-size effect leaves a signature of the tidal effect in the GW phase.
In the early stage of inspiral, the time scale for NS to restore hydrodynamical equilibrium ($T_{\mathrm{int}}$) is much shorter than the time scale for which the tidal field changes with orbital motion ($T_{\mathrm{ext}}$). These two time scales are approximately given by $T_{\mathrm{int}} \sim \left(G \rho \right)^{-1/2} \sim (R^3/Gm)^{1/2} $ and $T_{\mathrm{ext}} \sim (D^3/Gm)^{1/2}$ where $\rho, R$, and $m$ are the density, radius, mass of the NS, respectively and $D$ is the distance between the two stars. Since $R \ll D$, $T_{\mathrm{int}} \ll  T_{\mathrm{ext}}$  \cite{abdelsalhin2019tidaldeformationscompactobjects, Poisson_Will_2014}.
In particular, in this case, tidal effects are characterized by a single parameter called ``tidal deformability"  \cite{Dietrich:2020eud, PhysRevLett.116.181101, PhysRevD.77.021502}.
In the analysis of GW170817, the tidal deformability could be constrained for the first time, although not well measured   \cite{LIGOScientific:2017vwq, LIGOScientific:2018hze}.

The TaylorF2 waveform model is based on the post-Newtonian (pN) approximation
describing dynamics in the weak field regime.
Choi et al.  \cite{Choi:2018zbi} investigated the effect of 6 pN tidal correction on the measurement accuracy of the tidal deformability for non-spinning BNSs.
They showed that including 6 pN corrections can significantly reduce the measurement error of tidal deformability compared to using the leading order (5 pN) correction alone.
In the subsequent work, they extended the analysis to spinning BNSs and wider parameter space \cite{Choi:2022akz}.
On the other hand, higher-order corrections above 6 pN have been derived in many works, {\it e.g.}, 
spin-tidal coupling \cite{PhysRevD.98.104046}, tail effect \cite{PhysRevD.85.123007, PhysRevD.102.044033}, 
octupolar deformation \cite{PhysRevD.85.123007, PhysRevD.102.044033}, 
and spin-induced (rotational) tidal effect  \cite{PhysRevD.91.104018, PhysRevD.92.024010, PhysRevD.104.044052}.
The impact of the higher-order corrections on parameter estimation has also been studied by several authors  \cite{PhysRevD.98.124014, PhysRevD.98.124030, PhysRevD.106.024011, PhysRevD.106.103006, PhysRevD.108.063029}.
The tidal corrections are mainly contributed by gravito-electric tides that produce mass multipole moments,
and weakly by gravito-magnetic tides that induce the mass current multipoles.
The magnetic quadrupole tidal deformabilities are much smaller than the electric quadrupole tidal deformabilities by about $\mathcal{O}(0.01)$,
and are suppressed by 1-pN order in GW phase  \cite{PhysRevD.98.124014, PhysRevD.101.064003}.
Consequently, the leading-order (LO) contribution of electric tidal deformabilities first appears at 5 pN order,
while that of magnetic tidal deformabilities appears at 6 pN order.

In this work, we investigate how the measurement accuracy of NS tidal deformability can be improved in parameter estimation 
by incorporating the higher-order tidal corrections up to 7.5 pN including both the electric and magnetic contributions.
We assume a BNS system with spins aligned with orbital angular momentum and use the TaylorF2 waveform model, which is the most commonly used model
in parameter estimation for BNSs.
We calculate measurement errors for tidal deformability by using the Fisher matrix method.
Fisher matrix is an analytic approach used to predict the measurement errors of parameters  \cite{Vallisneri:2007ev}
and has been broadly used in GW data analysis studies.
We apply the Fisher matrix method to the TaylorF2 waveforms and show how much the accuracy of tidal deformability measurements improves.
Throughout this paper, we use geometrized unit $\left( G = c = 1 \right)$.


\section{ FISHER MATRIX AND TIDAL WAVEFORM}
\label{sec:GWandFM}
\subsection{Fisher Matrix Method}
In Bayesian inference,
the posterior probability that the GW signal in the data stream $s$ is characterized by the parameters $\theta$,
can be given by  the product of the prior $p(\theta)$ and likelihood $L(s \vert \theta)$,
\begin{equation}
    p ( \theta \vert s ) \propto p ( \theta) L(s \vert \theta).
\end{equation}
Assuming a uniform prior distribution, the posterior is only proportional to the likelihood.
If a GW signal is buried in the Gaussian noise, the likelihood can be represented as  \cite{Cutler:1994ys, Finn:1992wt}
\begin{equation}
    L\left( s \vert \theta \right) \propto \mathrm{exp} \left[ - \left\langle s-h \vert s-h \right\rangle /2 \right],
\end{equation}
where $h$ is a theoretical waveform and $\left\langle a \vert b \right\rangle$ 
is noise weighted inner product defined by  \cite{Cutler:1994ys, Finn:1992wt, Vallisneri:2007ev}

\begin{align}
    \left\langle a \vert b \right\rangle & \equiv 4 Re \int_0^{\infty} \frac{ \tilde{a}\left(f\right) \tilde{b}^{\ast}\left(f\right) }{S_n\left( f \right)} \ df, \label{inner product}
\end{align}
where $S_n (f)$ is power spectrum density (PSD), which represents the strength of noise depending on the frequency, 
and the asterisk in superscript denotes complex conjugate. 

If a signal-to-noise ratio (SNR) is sufficiently high, the likelihood exhibits a peak at the position of the true parameter value ($\theta_0$) and follows N-dimensional gaussian distribution expressed as $L \sim \mathrm{exp} \left[\ -\frac{1}{2} \Gamma_{ij} \Delta \theta_i \Delta \theta_j\ \right]$ \cite{Maggiore:2007ulw}
where $ \Delta\theta \equiv \theta - \theta_0 $.
Then, using the Fisher matrix defined by  \cite{Vallisneri:2007ev, Jaranowski:1994xd, PhysRevD.52.848, Maggiore:2007ulw}
\begin{equation}
    \Gamma_{ij} = \left\langle \frac{\partial h}{\partial \theta_i} \bigg\vert \frac{\partial h}{\partial \theta_j} \right\rangle \bigg\rvert_{\theta = \theta_0},
\end{equation}
one can obtain the covariance matrix using the relation $\Sigma = \Gamma^{-1} $.
The measurement error $\left( \sigma_i \right)$ and the correlation coefficient $\left( c_{ij} \right)$ between two parameters are given by
\begin{equation}
    \sigma_i = \sqrt{\Sigma_{ii}}, \ \ \ \ \ c_{ij} = \frac{\Sigma_{ij}}{\sqrt{\Sigma_{ii} \Sigma_{jj}}}.
\end{equation}
In general, the measurement error is inversely proportional to SNR.

In the Fisher matrix method, the prior information can be incorporated through the relation  \cite{Cutler:1994ys,PhysRevD.52.848}
\be
\Gamma^P_{ij}=\Gamma_{ij}+\Gamma_{ii}^0, \label{prior}
\ee
where $\Gamma^P_{ij}$ is the prior-incorporated Fisher matrix and $\Gamma_{ij}$ is the original Fisher matrix with no prior.
The component $\Gamma_{ii}^0$ is given by $\Gamma_{ii}^0=(P^2_{\theta_i})^{-1}$ where $P_{\theta_i}$ 
indicates the standard deviation of the Gaussian prior function on the parameter $\theta_i$.
On the right side of Eq. (\ref{prior}), since only $\Gamma_{ij}$ depends on the SNR,
the impact of prior on parameter measurement accuracy is stronger with lower SNRs  \cite{Cho_2022}.


\subsection{ Higher-order Tidal Waveforms}

The TaylorF2 waveform is expressed as
\begin{equation}
h(f) = A f^{-7/6} e^{i \Psi(f)}, \label{TaylorF2 waveform}
\end{equation}
where $A$ is an amplitude that only scales the signal strength, 
so it can be ignored in our analysis by selecting an appropriate SNR.
The phase is given by

\begin{equation}
    \begin{aligned} \label{waveform phase}
        \Psi\left( f \right) = 2\pi f  t_c  - \phi_c & - \frac{\pi}{4} + \Psi_{\mathrm{pp}} + \Psi_{\mathrm{spin}} + \Psi_{\mathrm{tidal}} 
    \end{aligned}
\end{equation}
where $t_c$ and $\phi_c$ are the coalescence time and coalescence phase, respectively.
$\Psi_{\mathrm{pp}}$ and $\Psi_{\mathrm{spin}}$ correspond to the contributions of the point particle and the spin, respectively.
In this work, we consider both $\Psi_{\mathrm{pp}}$ and $\Psi_{\mathrm{spin}}$ up to 3.5 PN  \cite{Buonanno:2009zt, PhysRevD.88.083002}.
$\Psi_{\mathrm{tidal}}$ indicates the phase corrections due to tidal effects.

The leading-order (LO) contribution of $\Psi_{\mathrm{tidal}}$ enters at 5 pN and depends only on the electric quadrupole tidal deformability $\Lambda_2$  \cite{LIGOScientific:2018hze, PhysRevD.77.021502}.
The higher-order corrections are recently computed up to 7.5 pN by the Effective Action formalism  \cite{PhysRevD.102.044033},
and the full expression of $\Psi_{\mathrm{tidal}}$ is given by 
\footnote{Note that the notation is different from Ref.  \cite{PhysRevD.102.044033}, but we have verified this equation is identical to the corresponding equation in Ref.  \cite{PhysRevD.102.044033}.}

\begin{widetext}
\be
    \Psi_{\mathrm{tidal}} =  \frac{3}{128\eta v^5}  \left[ -\frac{39}{2}  \tilde{\Lambda} v^{10}  
    +  \left( \frac{3115 \tilde{\Lambda}}{64} - \frac{6595\sqrt{1-4\eta} \  \delta \tilde{\Lambda}}{364} + \tilde{\Sigma} \right)  v^{12}  
    + ( \hat{K} + \hat{\Lambda} + \hat{\Sigma} ) v^{13}
    + ( \varepsilon \Lambda  + \delta \Sigma + \tilde{\Lambda}_3) v^{14}
    + ( \delta K + \hat{N} ) v^{15}    
    \right],  \label{phase AT}
\ee
\end{widetext}
where $v = \left[ \pi (m_1+m_2) f \right]^{1/3}$ is the post-Newtonian expansion parameter and $\eta=m_1 m_2 / (m_1+m_2)^2$ is the symmetric mass ratio.

The effective tidal parameters $\tilde{\Lambda}$ and $\delta \tilde{\Lambda}$ 
appeared at the LO and the next-to-leading-order (NLO, {\it i.e.} 6 pN), contributions of the electric quadrupole tidal deformability are defined by  \cite{PhysRevD.91.043002}
\begin{multline}  \label{LL1}
  \tilde{\Lambda} =  \frac{8}{13} \left[ \left( 1+7\eta -31\eta^2 \right) \left( \Lambda_{2,1} + \Lambda_{2,2} \right) \right. \\
  \left.+ \sqrt{1-4 \eta} \left( 1 + 9\eta - 11 \eta^2 \right) \left( \Lambda_{2,1} - \Lambda_{2,2} \right) \right],  \\
\end{multline}
\begin{multline}  \label{LL2}
    \delta \tilde{\Lambda} = \frac{1}{2} \left[ \sqrt{1-4\eta} \left( 1 - \frac{13272\eta}{1319} + \frac{8944\eta^2}{1319} \right) \left( \Lambda_{2,1} + \Lambda_{2,2} \right) \right.\\
    \left.+ \left( 1 - \frac{15910\eta}{1319} + \frac{32850\eta^2}{1319} + \frac{3380\eta^2}{1319}\right) \left( \Lambda_{2,1} - \Lambda_{2,2} \right) \right],
\end{multline}
where $ \Lambda_{2,i}$ is the electric quadrupole tidal deformability of $i$th NS.
The effective tidal parameters are much more efficient than using $ \Lambda_{2,1}$ and $ \Lambda_{2,2}$ and thus have been commonly used in past parameter
estimation studies.
$\tilde{\Sigma}$ is the LO contribution of magnetic quadrupole tidal deformability,
$\hat{K}$ is the LO electric type tidal-tail coupling, and
 $\hat{\Lambda}$ and $\hat{\Sigma}$  represent the electric and magnetic tidal-spin coupling effects, respectively.
The explicit expressions of these four terms are given in Ref.  \cite{PhysRevD.98.124014}.

In the 7 pN order, $\varepsilon \Lambda$ and $\delta \Sigma$ are the next-to-next-to-leading-order (NNLO) contribution of the electric quadrupole tidal deformability
and the NLO contribution of the magnetic quadrupole tidal deformability, respectively,
and $ \tilde{\Lambda}_3$ is the LO contribution of the electric octupolar tidal deformability.
In the 7.5 pN order, $\delta K$ and $\hat{N}$ are the NLO contribution of the electric tidal-tail coupling
and the LO contribution of the magnetic tidal-tail coupling, respectively.
In accordance with the notation in Ref.  \cite{PhysRevD.98.124014}, we present the explicit expressions of the five terms as follows.
\begin{multline}
    \varepsilon \Lambda = \left( \frac{72078065}{381024} - \frac{969 934 675}{571 536 X_1} + \frac{289 295 X_1}{378} \right. \\
    \left. + \frac{1 047 475 X_1^2}{1512} - \frac{4 825 X_1^3}{54} + 20 X_1^4 \right) X_1^5 \Lambda_{2,1} + ( 1 \leftrightarrow 2 ),
\end{multline}
\begin{multline}
     \delta \Sigma = -\left( \frac{3806540}{1701} - \frac{6101435}{1701 X_1} + \frac{64360 X_1}{81} \right. \\
     \left. + \frac{15080 X_1^2}{27} \right) X_1^5 \Sigma_{2,1} + ( 1 \leftrightarrow 2 ),
\end{multline}
\begin{equation}
    \tilde{\Lambda}_3 = \frac{4000}{9} \left( 1- \frac{1}{X_1} \right) \Lambda_{3, 1} + ( 1 \leftrightarrow 2 ), \\
\end{equation}
\begin{multline}
    \delta K = \pi \left( - \frac{22415}{28} + \frac{27719}{28X_1} + \frac{3799X_1}{14} \right. \\
    \left. - \frac{2630X_1^2}{7} \right) X_1^5 \Lambda_{2,1} + ( 1 \leftrightarrow 2 ),
\end{multline}
\begin{equation}
    \hat{N} = -\frac{16}{7} \left( 1 - \frac{2078}{21 X_1} \right) X_1^4 \Sigma_{2,1} + ( 1 \leftrightarrow 2 ),
\end{equation}
where $\Lambda_{3,i}$ and $\Sigma_{2,i}$ are the electric octupole and the magnetic quadrupole tidal deformabilities of $i$th NS, respectively,
and $X_i = m_i/(m_1+m_2)$.

\begin{table*}[t]
    \begin{ruledtabular}
        \begin{tabular}{cccccccc}
            Relation & $a_0$ & $a_1$ & $a_2$ & $a_3$ & $a_4$ & $a_5$ & $a_6$ \\ \hline\hline
            $\Lambda_3$ - $\Lambda_2$ & $-1.163$ & $9.442 \times 10^{-1}$ & $2.492 \times 10^{-1}$ & $-8.170 \times 10^{-2}$ & $1.374 \times 10^{-2}$ & $-1.117 \times 10^{-3}$ & $3.494 \times 10^{-5}$ \\
            $\Sigma_2$ - $\Lambda_2$ & -2.019 & $4.821 \times 10^{-1}$ & $7.609 \times 10^{-4}$ & $4.096 \times 10^{-3}$ & $-5.0 \times 10^{-4}$ & $2.643 \times 10^{-5}$ &$-5.192 \times 10^{-7}$  \\
        \end{tabular}
    \end{ruledtabular}
    \caption{Coefficients of the UR in Eq. \ref{UR}.  We use the most recent UR coefficients given in Ref.  \cite{PhysRevD.107.023010}.} 
    \label{UR_coefficient}
\end{table*}

For each NS, the above tidal terms consist of the three independent parameters, $ \Lambda_{2},  \Lambda_{3}$,
and $\Sigma_2$.
We here adopt Universal Relation (UR) to reduce the number of parameters.
UR is an approximate empirical formula connecting $\Lambda_3$ and $\Sigma_2$  to $\Lambda_2$,
which is not sensitive to the EOS model. 
This relation is useful to infer the physical quantities of NSs from the others even in the lack of information about EOS  \cite{doi:10.1126/science.1123430, universe5070159}.
Thus, UR allows us to reduce the parameter space dimension of the Fisher matrix.
UR can be expressed as
\begin{equation} \label{UR}
    \mathrm{ln}\, P = \sum_{k=0}^6 a_k \left( \mathrm{ln}\, \Lambda_2 \right)^{k},
\end{equation}
where $P$ denotes  $\Lambda_3$ or $\Sigma_2$.
The specific coefficients $a_k$ are given in Table \ref{UR_coefficient},
where we adopt the most recent version for $a_k$ given in Ref.  \cite{PhysRevD.107.023010}.
Using the UR, all terms in 
$\Psi_{\mathrm{tidal}} $ can be expressed by $\Lambda_{2,1}$ and $\Lambda_{2,2}$.
Finally, we convert ($\Lambda_{2,1}, \Lambda_{2,2}$) to ($\tilde{\Lambda}, \delta \tilde{\Lambda}$) using Eqs. (\ref{LL1}) and (\ref{LL2}).

We also use the chirp mass defined by $M_c = \left(m_1 m_2 \right)^{3/5}/(m_1+m_2)^{1/5}$ and the symmetric mass ratio rather than the two individual masses.
In addition, since the two spins are strongly correlated in align-spin systems, it is useful to use a single effective spin parameter
defined by the mass-weighted linear combination as   \cite{PhysRevD.82.064016, PhysRevLett.106.241101},
\begin{equation}
    \chi_{\mathrm{eff}} = \frac{m_1 \chi_1 + m_2 \chi_2}{m_1 + m_2}.
\end{equation}
Consequently, our Fisher matrix is given by a 7 $\times$ 7 matrix with the parameters, $\theta_i = \left\{M_c, \eta, \chi_{\mathrm{eff}},  \tilde{\Lambda}, \delta \tilde{\Lambda}, t_c, \phi_c \right\}$.

\section{RESULT}
\label{sec:result}
We calculate the measurement errors of parameters for BNSs and investigate how much the accuracy in measuring tidal deformability $\tilde{\Lambda}$
can be improved by incorporating the higher-order tidal corrections.
Referring to Ref.  \cite{PhysRevD.105.124022}, we apply the prior $P_{\delta \tilde{\Lambda}}=500$ in our calculation for efficiency.
Here we assume a fiducial BNS system with physical parameters $m_1= 1.6M_{\odot}$, $m_2=1.2M_{\odot}, \chi_{\rm eff}=0.05, \tilde{\Lambda}=266.7$, and $\delta \tilde{\Lambda}=56.9$
that are roughly compatible with GW170817.
The true values of $\tilde{\Lambda}$ and $\delta \tilde{\Lambda}$  are obtained according to APR4
 \cite{PhysRevC.58.1804, PhysRevD.95.063016}, which is one of the EOS models well describing the result of GW170817.
We use the zero-detuned, high-power noise PSD for aLIGO  \cite{aLIGOpsd} and set the integration interval in Eq. (\ref{inner product}) from 10 to 1000 Hz.

\subsection{Tidal Dephasing Due to Higher-order Corrections}

\begin{figure}[t]
    \centering 
    \includegraphics[width=1\columnwidth]{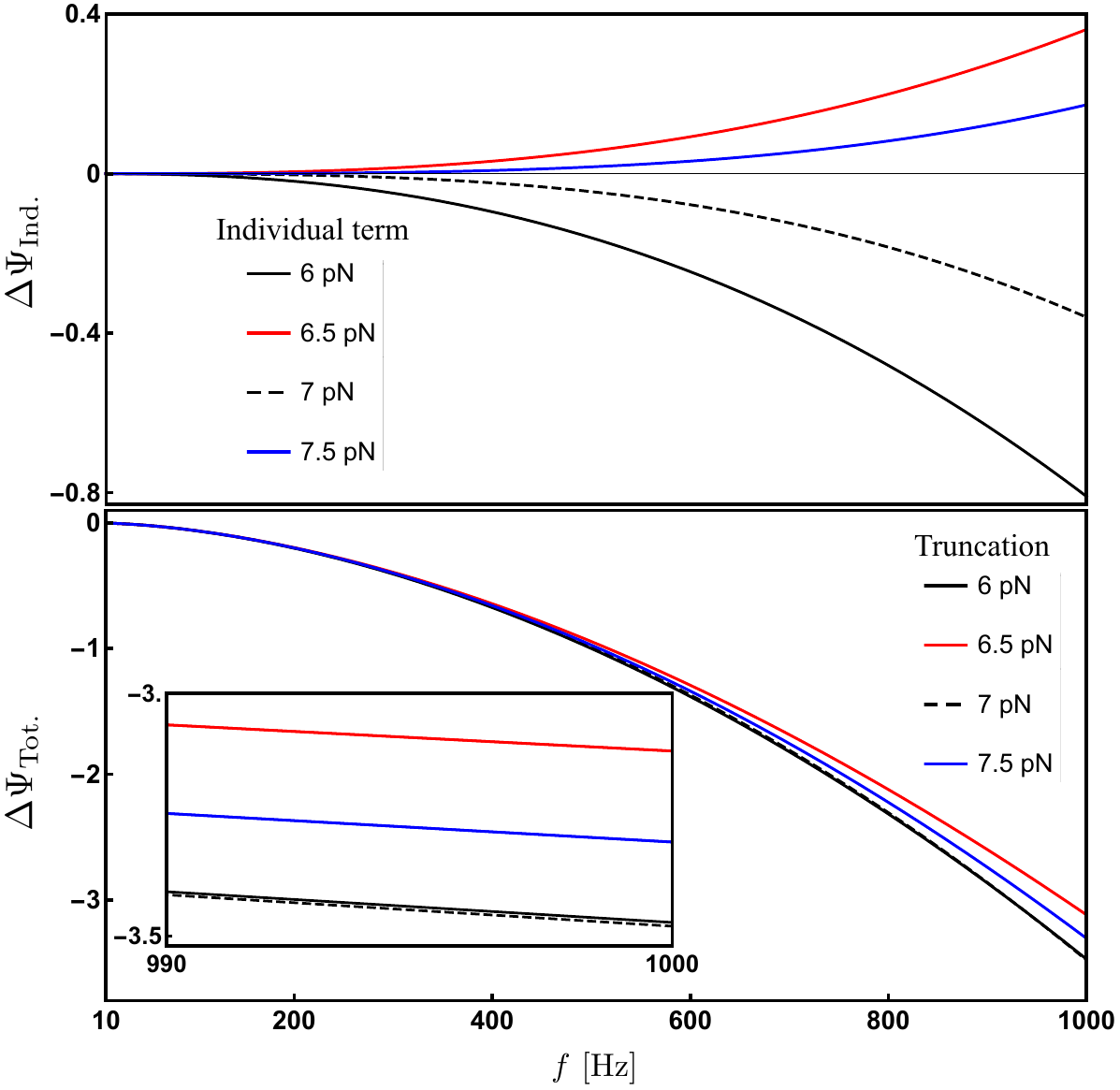}
    \caption{Cumulative tidal dephasing ($\Delta \Psi$) generated by individual tidal correction terms (top panel) and by tidal corrections truncated at the n pN order (bottom panel). We assume a BNS system with physical parameters $m_1= 1.6M_{\odot}$, $m_2=1.2M_{\odot}, \chi_{\rm eff}=0.05, \tilde{\Lambda}=266.7$, and $\delta \tilde{\Lambda}=56.9$
that are roughly compatible with GW170817.
The true values of the tidal parameters are obtained according to APR4
 \cite{PhysRevC.58.1804, PhysRevD.95.063016}, which is one of the EOS models well describing the result of GW170817.
Note that as the pN order increases, the sign of $\Delta \Psi_{\rm Ind}$ alternates and the convergence is nonmonotonic.}
     \label{dephasing}
\end{figure}

Fig. \ref{dephasing} shows the evolution of tidal dephasing ($\Delta \Psi$)
between $f=10$ and 1000 Hz.
We consider the tidal terms individually from 6 pN to 7.5 pN given in Eq. (\ref{phase AT}).
The top panel shows the cumulative tidal dephasing generated by the tidal terms individually from 6 pN to 7.5 pN given in Eq. (\ref{phase AT}).
For reference, in the case of 5 pN (not shown in this result), $\Delta \Psi_{\rm Ind.} \simeq -2.661$ at 1000 Hz.
For 6 pN (black line), $\Delta \Psi_{\rm Ind.} \simeq-0.799$,
and $\Delta \Psi_{\rm Ind.} \simeq  0.352, -0.354,$ and 0.163 for 6.5 pN (red line), 7 pN (dashed line), and 7.5 pN (blue line), respectively.
In the bottom panel, we describe the total tidal dephasing $\Delta \Psi_{\rm Tot.}$ generated by the tidal corrections truncated at the n pN order ({\it i.e.},  all tidal corrections from 5 pN to n pN).
The black line is obtained from the 6 pN-truncated corrections,
where $\Delta \Psi_{\rm Tot.} \simeq -3.460$ at 1000 Hz.
For 6.5 pN (red line), 7 pN (dashed line), and 7.5 pN  (blue line)-truncated corrections, $\Delta \Psi_{\rm Tot.} \simeq -3.107, -3.461$, and  $-3.298$, respectively.

We find that as the truncation pN order increases,  the convergence is nonmonotonic
because the sign of $\Delta \Psi_{\rm Ind.}$ alternates (top panel in Fig. \ref{dephasing}).
This behavior was also demonstrated in Ref.  \cite{PhysRevD.85.123007} (refer to Fig. 1 therein).
In addition, Forteza \etal  \cite{PhysRevD.98.124014} showed that including 6.5 pN term can decrease the accuracy of the TaylorF2 tidal waveform
compared to 6 pN-truncated corrections.
We find that the opposite dephasing effect due to the 6.5 pN term 
is mainly attributed to the LO tidal-tail coupling ($\hat{K}$) and is weakly dependent on the tidal-spin coupling ($\hat{\Lambda}$).
Similarly, the opposite dephasing effect due to the 7.5 pN term is caused by the NLO tidal-tail coupling ($\delta K$).

\begin{table}[t]
    \begin{ruledtabular}
        \begin{tabular}{cccc}
           Truncation &   $(1.8, 1.0)M_{\odot}$  &  $(1.6, 1.2)M_{\odot}$ & $(1.4, 1.4)M_{\odot}$  \\ \hline\hline
        6 pN       &     303.3 (0.937)   &317.7   (0.937)                     &322.5 (0.937) \\ 
       6.5 pN      &     362.3 (1.120)     &380.7   (1.123)                    &387.0 (1.124)\\ 
        7 pN         &    296.0 (0.915)  &309.1      (0.912)                &313.7 (0.911) \\ 
      7.5 pN        &     323.6  (1.000) &338.9    (1.000)                    &344.3 (1.000) \\   \hline
    $\tilde{\Lambda}_{\rm true}$ &  322.6   & 266.7&   250.0   \\
        \end{tabular}
    \end{ruledtabular}
    \caption{Measurement error of tidal deformability ($\sigma_{\tilde{\Lambda}}$) obtained by using the TaylorF2 waveform 
    where the tidal phase ($\Psi_{\rm tidal}$) is truncated at n pN order.
  The number in parenthesis indicates the ratio to the 7.5 pN-truncated result.  We set SNR to 32.4 consistent with GW170817.
  Note that $\sigma_{\tilde{\Lambda}}$ is smaller for larger value of $|\Delta \Psi_{\rm Tot.}|$.}
    \label{errors}
\end{table}

Incorporating the n pN-truncated tidal phase ($\Psi_{\rm tidal}$) in the TaylorF2 waveforms given in Eqs. (\ref{TaylorF2 waveform}) and (\ref{waveform phase}), 
we calculate measurement errors of $\tilde{\Lambda}$ for our fiducial binary 
and list the results in Table \ref{errors}.
We find that the measurement accuracy depends directly on the accumulated $\Delta \Psi_{\rm Tot.}$.
Roughly speaking, larger tidal dephasing can provide more information about the tidal parameter,
resulting in improved measurement accuracy.
Compared with the case of 7.5 pN, 6.5 pN overestimates the measurement error because it has less tidal dephasing,
while 6 pN and 7 pN underestimate due to larger tidal dephasing.
For comparison, we also show results for asymmetric and symmetric binaries with the same total mass as our fiducial binary.
At the same pN order level, the measurement accuracy is better for more asymmetric binaries.
Interestingly, the error ratios to the 7.5 pN error (given in parentheses) are almost identical between the three binaries, regardless of the mass ratios.

\subsection{Dependence on Spin $\&$ EOS}

\begin{figure}[t]
    \centering
    \includegraphics[width=1\columnwidth]{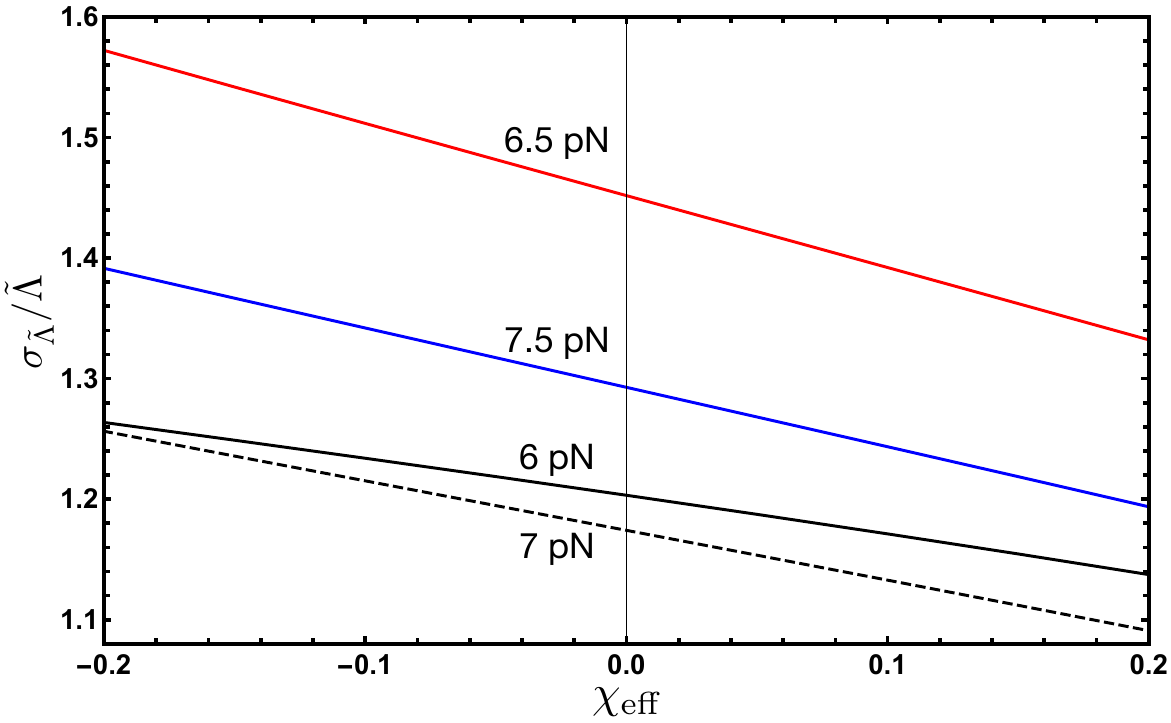}
    \caption{Dependence of the measurement error on the effective spin. 
    We assume our fiducial binary with $m_1= 1.6M_{\odot}$, $m_2=1.2M_{\odot}$, and $\tilde{\Lambda}=266.7$,
    and set SNR to 32.4.}
    \label{spin_error}
\end{figure}

Fig. \ref{spin_error} shows the change in measurement error ($\sigma_{\tilde{\Lambda}}/\tilde{\Lambda}$) according to the effective spin.
For all pN orders, the errors exhibit monotonically decreasing behavior with increasing spin from $-0.2$ to 0.2,
and this behavior is consistent over the entire physical spin range (see, Fig. 1 of Ref.  \cite{Choi:2022akz}).
This can be explained from the relationship between spin and GW phase ($\Psi$).
In aligned-spin binary systems, positive (negative) spins make phase evolution slightly slower (faster),
allowing them to accumulate larger (smaller) phases compared to the nonspinning case  \cite{PhysRevD.94.124045}.
In addition, the tidal-spin coupling ($\hat{\Lambda}, \hat{\Sigma}$) that appears at 6.5 pN is likely to make this effect stronger,
and thus the slope for the three higher-order results is steeper.

\begin{figure}[t]
    \centering
    \includegraphics[width=1\columnwidth]{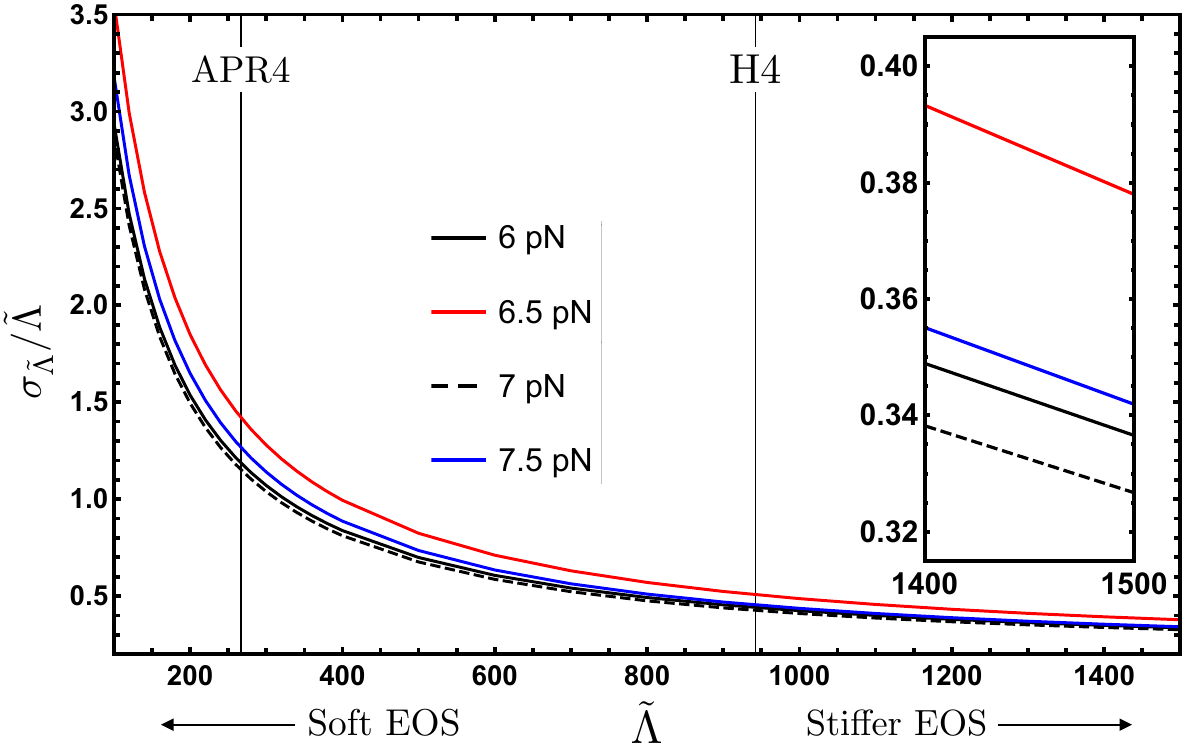}
    \caption{Measurement error ($\sigma_{\tilde{\Lambda}}/\tilde{\Lambda}$) obtained by varying the tidal deformability.
    We assume our fiducial binary with $m_1= 1.6M_{\odot}$, $m_2=1.2M_{\odot}$, and $\chi_{\mathrm{eff}} = 0.05$,
    and set SNR to 32.4.
    The vertical lines denote the values of $\tilde{\Lambda}$ obtained from the APR4 (left) and H4 (right) models.}
    \label{tidal}
\end{figure}

In Fig. \ref{tidal}, we calculate the measurement error ($\sigma_{\tilde{\Lambda}}/\tilde{\Lambda}$) by varying the tidal deformability.
Here, we assume the same binary masses as in Fig. \ref{spin_error} with a fixed spin value of $\chi_{\rm eff}=0.05$,
but vary the true value of $\tilde{\Lambda}$ from 100 to 1500.\footnote{In this calculation, true values of $\delta \tilde{\Lambda}$ 
were determined by using a quadratic fitting function which increases monotonically from $\sim 20$ to $\sim 240$ in the range of $100 \leq \tilde{\Lambda} \leq 1500$.
This function was obtained from several EOS models. }
In this plot, the low (larger)-$\tilde{\Lambda}$ region corresponds to a soft (stiffer) EOS model,
and we display the soft EOS model APR4 and the stiffer model H4 for reference.
We find that the fractional error ($\sigma_{\tilde{\Lambda}}/\tilde{\Lambda}$) decreases rapidly in the low-$\tilde{\Lambda}$ region,
and this tendency gradually mitigates with increasing $\tilde{\Lambda}$.
For the case of 7.5 pN, we obtain $\sigma_{\tilde{\Lambda}}/\tilde{\Lambda} \sim 1.3$ and 0.5 for APR4 and H4 models, respectively.
In parameter estimation of GW170817  \cite{LIGOScientific:2018hze}, the symmetric 90$\%$ credible interval for $\tilde{\Lambda}$ with Low-spin prior
is given as $340^{+580}_{-240}$, which is roughly consistent with our results for 6 pN. \footnote{In Fig.  \ref{tidal}, for 6 pN, $\sigma_{\tilde{\lambda}} \simeq  327$ when $\tilde{\lambda}=340$,
which corresponds to 90\% credible interval of $340^{+538}_{-538}$.}

\section{SUMMARY AND DISCUSSION}
\label{sec:conclusions}

In this work, we investigated the impact of higher-order tidal corrections on the measurement of NS tidal deformability.
We used a recent TaylorF2 waveform model, which incorporated tidal corrections up to 7.5 pN, including electric and magnetic tidal effects.
Employing UR, we removed additional parameters that appeared in the higher-order corrections beyond 6 pN. 
Investigating the effect of individual pN-order tidal corrections from 6 pN to 7.5 pN on the GW phase,
we found different signs of dephasing ($\Delta \Psi_{\rm Ind.}$) between pN orders,
resulting in a nonmonotonic convergence of the total tidal dephasing ($\Delta \Psi_{\rm Tot.}$) with increasing truncation pN orders.
We chose a binary with physical parameters ($m_1=1.6 \msun, m_2=1.2 \msun, \chi_{\rm eff}=0.05, \tilde{\lambda}=266.7, \delta \tilde{\Lambda}=56.9, {\rm SNR}=32.4$) 
compatible with GW170817 as our fiducial BNS system.
We calculated measurement errors of tidal deformability using higher-order tidal corrections.
As in the case of the tidal dephasing, increasing the pN order did not improve the measurement accuracy.
Compared with the case of 7.5 pN, 6.5 pN has less total tidal dephasing and a larger measurement error,
while 6 pN and 7 pN have larger total tidal dephasings and smaller measurement errors.
We investigated the change in measurement error by varying the true value of $\chi_{\rm eff}$ or  $\tilde{\lambda}$ with all the other parameters fixed.
For all pN orders, the errors monotonically decrease as $\chi_{\rm eff}$ increases due to the effect of phase delay.
Finally, we found that the tidal deformability can be better measured for a stiffer EOS.
For the case of 7.5 pN,  $\sigma_{\tilde{\Lambda}}/\tilde{\Lambda} \sim 1.3$ and 0.5 for APR4 and H4 models, respectively.

Given the nonconvergent behavior described in this work, it would be unnecessary to develop higher-order tidal corrections beyond 7.5 pN.
It has recently been shown that dynamical tides (DT) can produce tidal dephasing comparable to that of the adiabatic tides considered in this work. 
When the frequency of the gravitational driving force approaches the NS mode frequency, DT can induce the normal mode resonance.
In particular, the fundamental mode is the dominant contribution among several normal modes \cite{10.1093/mnras/275.2.301, 10.1143/ptp/91.5.871, 10.1093/mnras/stab870, PhysRevD.101.083001}
and is expected to be measurable with the 3-G detectors \cite{PhysRevD.105.123032}.
Ignoring the DT effects in parameter estimation can produce systematic bias in tidal deformability measurements,
resulting in inaccurate predictions for NS EOS\cite{PhysRevLett.129.081102, Pradhan:2023xtq, PhysRevD.107.023010}.
We will extend our analysis to include the DT effect in future works.

\section*{Acknowledgments}
This work was supported by 2024 BK21 FOUR Graduate School Innovation Support funded by Pusan National University (PNU-Fellowship program),
and the National Research Foundation of Korea (NRF) grants funded by the Korea government 
(No. RS-2023-NR076639, No. RS-2023-00242247, and No. RS-2023-00301938).

\bibliographystyle{aip}
\bibliography{biblio}

\end{document}